\documentclass[pdflatex,sn-mathphys-num]{sn-jnl}

\usepackage{graphicx}%
\usepackage{multirow}%
\usepackage{amsmath,amssymb,amsfonts}%
\usepackage{amsthm}%
\usepackage{mathrsfs}%
\usepackage[title]{appendix}%
\usepackage{xcolor}%
\usepackage{textcomp}%
\usepackage{manyfoot}%
\usepackage{booktabs}%
\usepackage{algorithm}%
\usepackage{algorithmicx}%
\usepackage{algpseudocode}%
\usepackage{listings}%

\theoremstyle{thmstyleone}%
\theoremstyle{thmstyletwo}%

\theoremstyle{thmstylethree}%

\begin{document}

\title[Article Title]{Dynamic Sampling for Telemetry in Microservices: A Reinforcement Learning and Entropy-Based Approach}


\author[1]{\fnm{Renan Martins} \sur{Alves}}\email{rmalves@inf.ufrgs.br}

\author[1]{\fnm{Jéferson Campos} \sur{Nobre}}\email{jcnobre@inf.ufrgs.br}

\author*[1]{\fnm{Juliano} \sur{Araujo Wickboldt}}\email{jwickboldt@inf.ufrgs.br}

\affil*[1]{\orgdiv{Federal University of Rio Grande do Sul}, \orgname{UFRGS}, \orgaddress{\street{Av. Bento Gonçalves 9500}, \city{Porto Alegre}, \postcode{90046-900}, \state{RS}, \country{Brazil}}}


\abstract{Microservices architectures are increasingly deployed in cloud-based distributed environments, making application development and maintenance more dynamic, but also increasing the complexity of troubleshooting and observability. Distributed tracing tools are therefore essential for request analysis and debugging, despite introducing additional overhead that can be amplified by excessive and inefficient data collection. This article proposes RADAR (Reinforcement Learning Agent for Dynamic And Relevant trace sampling), an agent that combines reinforcement learning with a data entropy assessment to achieve more efficient capture of traces relevant to system monitoring, based on the OpenTelemetry standard. RADAR tests different sampling rules to discover which combination is most efficient. A test environment simulating a minimalist online store with several microservices distributed across a Kubernetes cluster served as the basis for the experiments, which evaluated the agent's convergence and the system's performance in terms of resource consumption and collected data quality. Results showed that RADAR reduced network bandwidth consumption by 97.4\% and CPU usage by 99.0\% compared to full data collection, also outperforming a fixed-rate sampling baseline. Beyond these resource savings, the approach preserved observability of critical scenarios, retaining approximately 85.6\% of rare trace patterns and increasing the average entropy of the stored information by approximately 25\%, validating the feasibility of using entropy to orchestrate telemetry autonomously and efficiently.}




\maketitle

\section{Introduction}\label{intro}

The transition from monolithic architectures to microservices has allowed organizations to increase development agility and application scalability~\cite{kratzke2018brief}. However, this transition has also increased the complexity of performing system observability tasks~\cite{zhou2018fault}. In microservices environments, a single user request may traverse dozens of independent services, making distributed tracing an indispensable tool for understanding execution flow and diagnosing latent failures~\cite{zhou2018fault}. OpenTelemetry has emerged as an open standard that enables the generation, collection, and export of telemetry data across a wide range of languages and platforms, allowing end-to-end request tracing.

Despite its importance, collecting traces in full at large scale is infeasible: the resulting data volume can overwhelm network bandwidth, consume excessive CPU cycles for processing, and lead to prohibitive storage costs~\cite{hindsight}. To mitigate this problem, sampling strategies are applied. Traditional methods, such as probabilistic head-sampling, decide whether to discard data at the beginning of a transaction, failing to capture rare events or errors that only manifest later in the flow. Tail-sampling, in turn, allows for more informed decisions, but its manual configuration is rigid and unable to adapt to the dynamic load and behavior variations typical of cloud environments~\cite{blanco2023practical}. Robust solutions such as Hindsight~\cite{hindsight}, which generates and retains all traces locally and only forwards them to the collector once an anomaly is detected, or the work of Las-Casas et al.~\cite{lascasas}, which clusters similar and common traces into tree nodes and prioritizes their discard when resources need to be freed, still leave notable gaps, such as rigid trigger definitions for anomalies or a lack of compatibility with more modern observability tooling.

The decision of which traces to retain depends on (i) the dynamic characteristics of traffic, (ii) the occurrence of errors and anomalies, and (iii) the diversity and informational richness of the executed paths. This decision must be made continuously, since system behavior changes over time; static, manually defined sampling heuristics are unable to adapt efficiently to complex and variable scenarios. This leads to the central challenge addressed in this work: \textbf{\textit{how to significantly reduce the volume of traces collected in distributed systems while preserving the high informational value required for observability?}}

In this scenario, there is a need for observability mechanisms that can autonomously select the most relevant data. This work proposes \textbf{RADAR} (Reinforcement Learning Agent for Dynamic And Relevant trace sampling), a framework that uses Reinforcement Learning (RL) to dynamically adjust sampling policies in OpenTelemetry collectors. RADAR uses \textbf{Shannon entropy} as its reward metric, allowing the system to prioritize traces that present greater informational diversity, while discarding redundant requests that add no diagnostic value.

The main objective of this article is to present a dynamic sampling architecture capable of selecting, from a catalog of tail-sampling rules, the combination that preserves the traces with the highest diversity and diagnostic utility, balancing operational cost against information quality, without relying on statistical heuristics or prior data labeling. The main contributions of this work include:
\begin{itemize}
    \item \textbf{The RADAR framework}: the design and practical implementation of an RL agent integrated into the OpenTelemetry ecosystem, capable of adjusting sampling policies at runtime;
    \item \textbf{An entropy-based reward metric}: a methodology for quantifying the relevance of distributed traces, balancing operational cost and information value, without requiring prior data labeling;
    \item \textbf{An experimental microservices environment}: a test environment with controllable traffic and fault injection, used to evaluate the agent's convergence and the system's performance.
\end{itemize}

The remainder of this article is organized as follows. Section~\ref{fundamentals} presents the theoretical background on observability, OpenTelemetry, Shannon entropy, and reinforcement learning. Section~\ref{relatedwork} discusses related work on sampling strategies and telemetry optimization. Section~\ref{proposal} describes the proposed methodology, system architecture, and problem modeling. Section~\ref{prototype} details the RADAR prototype implementation. Section~\ref{evaluation} presents the experimental evaluation and results. Finally, Section~\ref{conclusion} presents concluding remarks and directions for future work.

\section{Fundamentals}\label{fundamentals}

Distributed systems composed of independent, communicating components have become the standard architecture for modern cloud applications, with microservices further decomposing these systems into small, independently deployable services that communicate over the network~\cite{distributedSystems, Dragoni2017}. Containerization, popularized by Docker, and orchestration platforms such as Kubernetes are the de facto infrastructure for deploying and scaling such systems~\cite{docker_article, kubernetes_up_and_running}.

Observability allows the internal state of a complex system to be inferred from data exposed externally, overcoming the limitations of traditional metrics and logs in highly dynamic microservices environments~\cite{ObservEng, zhou2018fault}. Solving this problem requires context propagation and automatic correlation between events; the system must be natively instrumented to emit traces, logs, and metrics, ensuring the information needed for diagnosis is available without further intervention~\cite{ObservEng}. A widely adopted concept in modern observability is the structured event: a record of everything that happened during a service's execution, organized as a map of keys for easy data access. A set of related structured events forms a distributed trace, offering a continuous, detailed view of a request's flow from start to end, rather than isolated metrics~\cite{ObservEng}.

OpenTelemetry has established itself as the standardized framework for collecting telemetry data from complex systems, offering tools for logging, metrics, and tracing across a wide range of programming languages~\cite{ObservEng}. Its two central components are traces, the record of the complete path taken by a request, and spans, the individual units of work that compose a trace, together providing detailed context for each step of execution in a distributed system~\cite{ObservEng}. Telemetry data is typically collected, processed, and exported by an OpenTelemetry Collector, a pipeline composed of receivers (which ingest data), processors (which filter, transform, or sample it), and exporters (which forward it to one or more observability backends such as Jaeger~\cite{jaeger_intro}).

To optimize processing and storage costs without compromising observability, sampling techniques are employed~\cite{blanco2023practical}. Head-based sampling decides whether to collect a trace at the very start of a request, which is simple and inexpensive but lacks the context of the complete request. Tail-based sampling, in contrast, defers the decision until after a trace concludes, enabling more granular and sophisticated criteria, such as always retaining traces with errors or abnormal latency, at the cost of buffering all of a trace's spans until the decision is made~\cite{ObservEng, blanco2023practical}.

Shannon entropy is a central concept in information theory, introduced by Claude E. Shannon in his classic 1948 paper \textit{A Mathematical Theory of Communication}. Shannon formalized communication as the transmission of messages through possibly noisy channels and proposed a quantitative measure of the uncertainty associated with a source of symbols, which he termed entropy~\cite{shannon1948mathematical}. Formally, consider a discrete random variable $X$ taking values in a finite alphabet $\{x_1, x_2, \ldots, x_n\}$ with probabilities $P(x_i)$. Shannon entropy is defined as~\cite{cover2006elements}:
\begin{equation}
    H(X) = -\sum_{i=1}^{n} P(x_i) \log_{b} P(x_i)
    \label{eq:shannon}
\end{equation}
where $b$ is the base of the logarithm (commonly $b=2$). Intuitively, this expression represents the average amount of information per symbol emitted by the source: rare events contribute more information (a larger $-\log P(x_i)$), while highly probable events contribute less. Entropy is always non-negative, equals zero only when the source is deterministic, and reaches its maximum value $\log_b n$ when all symbols are equally likely, characterizing the state of maximum uncertainty. Beyond telecommunications, Shannon entropy is widely used in statistics and machine learning as a measure of disorder, spread, or uncertainty in complex systems~\cite{mackay2003information}.

Reinforcement learning (RL) is a machine learning approach in which an agent learns to make decisions through trial and error, interacting with an environment and receiving reward signals in response to its actions. Unlike supervised learning, where a model is given correct labels for each example, in RL the agent receives only a scalar feedback signal (the reward) and must discover, from that signal alone, a decision policy that maximizes the cumulative return over time~\cite{sutton2018reinforcement}.

The classical RL setting is formalized as a Markov Decision Process (MDP), described by a set of states, a set of actions, a state-transition function, and a reward function. At each time step $t$, the agent observes a state $s_t$, selects an action $a_t$ according to a policy $\pi(a|s)$, and the environment responds with a scalar reward $r_{t+1}$ and a new state $s_{t+1}$; the agent's objective is to maximize the expected return, typically the discounted sum of future rewards. This scheme highlights the three central components of RL: the agent (who decides), the environment (with which it interacts), and the reward (the signal that guides learning)~\cite{sutton2018reinforcement}. The agent implements the decision policy, mapping observations to actions, and must balance exploration (trying new actions to gather information) and exploitation (leveraging knowledge already acquired). The reward function is critical to this process: a poorly designed reward can lead the agent to undesirable behavior, even if that behavior is mathematically ``optimal'' for the specified metric~\cite{sutton2018reinforcement}.

A relevant special case arises when the environment has no meaningful observable state, i.e., the outcome of an action does not depend on any state that changes between decisions. This setting is known as a Multi-Armed Bandit, a stateless simplification of the MDP in which the agent repeatedly chooses among a fixed set of actions (``arms'') to maximize cumulative reward, relying solely on the history of rewards obtained rather than on state transitions~\cite{sutton2018reinforcement}. This formulation is particularly suited to problems where decisions are made independently at each round, without requiring the agent to track or infer an evolving system state.

Reinforcement learning agents such as RADAR's are increasingly used to realize \textit{autonomic} management loops, in which a system monitors itself, analyzes and plans corrective or adaptive actions, and executes them with minimal human intervention. This pattern is commonly described by the MAPE-K reference model~\cite{kephart2003vision}: a \textbf{Monitor} stage collects data about the managed system, an \textbf{Analyze} stage derives relevant metrics or symptoms, a \textbf{Plan} stage decides on an adaptation, and an \textbf{Execute} stage applies it, with a shared \textbf{Knowledge} base persisting across cycles. Recent network and service management research applies this pattern with RL agents that periodically reconfigure a system to meet a management objective -- for example, dynamically tuning routing, traffic blocking, and scaling to satisfy end-to-end performance objectives on a service mesh~\cite{samani2024servicemesh}, or combining attention-based workload forecasting with policy-gradient RL for SLA-aware autoscaling in edge-cloud environments~\cite{shaikh2025autoscaling}.

\section{Related Work}\label{relatedwork}

Beyond adaptive sampling strategies, distributed tracing has increasingly become a first-class data source for network and service management research. Recent work uses trace data for root-cause localization of microservice anomalies, aggregating invocations at the operation level and ranking candidate root causes with a personalized PageRank algorithm~\cite{yang2024micronet}, and for anomaly detection via a dual autoencoder that jointly models the structural and temporal properties of service invocation graphs~\cite{li2025tracedae}. These works illustrate the growing role of tracing infrastructure in network and service management, but both assume traces have already been fully collected; RADAR instead addresses the complementary, upstream problem of deciding which traces are worth collecting in the first place.

Las-Casas et al. propose increasing the probability of collecting unique traces by computing the Euclidean distance between traces represented as numerical vectors derived from event graphs~\cite{lascasas}. Using this distance metric, traces are organized via hierarchical clustering into a tree, with frequent events forming deep branches and rare traces remaining close to the root. Sampling is performed by having the algorithm walk randomly from the root to a leaf, which favors selecting the rare traces near the root. Although this approach exploits the structural diversity of the data, it depends on static clustering algorithms and a custom vector representation. In contrast, this work replaces the fixed heuristic with a reinforcement learning agent that adapts its policy dynamically, operating directly on standard OpenTelemetry infrastructure without requiring complex data conversions.

Poghosyan et al. initially combine head- and tail-sampling to reduce the volume of data passed to their proposed solution~\cite{xai}. The remaining data then undergoes a noise-reduction step in which rare traces are discarded and common ones preserved, since recurring errors are more likely to explain a systemic failure; the data is subsequently converted into tabular form and fed into a machine learning system (RIPPER), which generates simple rules pointing to failure conditions, each weighted by Dempster-Shafer theory to measure its uncertainty. In contrast to this noise-reduction step, which actively discards rare events, the approach proposed here seeks to maximize entropy specifically to capture rare and anomalous cases; it also replaces static rule generation with a continuous, reward-guided feedback loop, eliminating the need to convert hierarchical traces into flat tables for processing.

Luo et al. propose a blame-proportional logging system that uses lightweight triggers to detect the first occurrence of a problem at low cost, then assigns a blame ranking to the application's methods, indicating each one's likelihood of being relevant to the root cause~\cite{audit}. Once a problem is detected, the system activates heavy logging only on the methods most likely to be responsible, combining continuous low-cost monitoring, an analysis of the request's call graph, and specific criteria for exceptions or performance issues. This strategy concentrates aggressive data collection only where it is needed. In contrast, while this approach relies on manually defined declarative triggers and dynamic runtime instrumentation, the present work proposes a partially autonomous solution based on multiple rules and on entropy as a universal signal of interest, acting solely on the telemetry collector's sampling configuration.

Sharma and Nadig propose using kernel-level observability via eBPF for distributed systems~\cite{ebpf-enhanced}. eBPF agents are deployed on every node of the cluster, collecting detailed system metrics directly from the operating system kernel and exposing tracing points that are not accessible from user space; distributed tracing is then performed using the eBPF library Deepflow, which natively parses kernel network packets to extract tracing headers, producing complete distributed traces without requiring a user-space proxy. While this approach provides deep visibility, it requires privileged kernel access, which makes it unsuitable for many managed container environments. This work differs by operating entirely in user space through OpenTelemetry, ensuring greater portability, and by adding an active intelligence layer to optimize data volume, whereas the eBPF-based solution focuses primarily on passive metrics collection.

Targeting a different domain, Luo et al. propose Hubble, which addresses method tracing for rare, intermittent problems on Android devices, where production overhead is a critical constraint~\cite{hubble}. Hubble uses just-in-time tracing to continuously record tracing data in a circular in-memory queue, where older data is constantly overwritten; this queue is only persisted to disk when the system's anomaly detector fires, preserving the crucial execution history that led to the problem. To achieve near-zero performance overhead, Hubble relies on aggressive low-level optimizations, inserting tracing logic directly into each method's compiled code through modifications to the Android compiler, with the most performance-critical code hand-written in assembly. The key distinction lies in domain and applicability: Hubble is a highly specialized solution for mobile devices that requires invasive compiler and hand-written assembly modifications, whereas the present work targets microservice orchestration in cloud backends, prioritizing compatibility with industry standards and language-agnostic instrumentation, without requiring low-level component rewrites.

Zhang et al. propose Hindsight, a distributed tracing system designed to efficiently and reliably capture edge cases in large-scale environments~\cite{hindsight}. Hindsight is based on retroactive sampling: all spans are generated locally, but only collected and forwarded to the backend once a symptom of anomaly is detected, through programmable triggers that identify, for example, elevated latency, errors, or anomalous behavior. Hindsight retroactively recovers the data belonging to the problematic request by querying distributed agents that temporarily buffer spans in memory. The authors show that, by decoupling trace generation from ingestion, the system captures nearly all rare cases with minimal impact on application performance, operating with nanosecond-scale latency per generated event, and that it is compatible with existing APIs such as OpenTelemetry and X-Trace. Despite this compatibility, Hindsight operates reactively, depending on predefined anomaly triggers to decide what data to preserve; the present work advances this by using reinforcement learning to proactively discover which traces are relevant through entropy maximization, capturing scenarios of interest that would not activate a conventional static trigger.

Table~\ref{tab:related-work-comparison} summarizes these related approaches alongside RADAR, highlighting the conceptual and technical differences in sampling and observability strategy. Unlike the existing solutions, RADAR combines reinforcement learning with entropy maximization to perform dynamic adjustments without relying on manual triggers or invasive instrumentation. Its native compatibility with OpenTelemetry further distinguishes it from approaches based on static heuristics, predefined rules, or reactive mechanisms, reinforcing its original contribution in the context of adaptive telemetry sampling.

\begin{table}[ht]
\centering
\caption{Comparison between related work and RADAR}
\label{tab:related-work-comparison}
\begin{tabular}{lccccc}
\toprule
Related & RL    & Prioritizes  & Dynamic   & OTel          & Microservices     \\
Work    &       & rare traces  & Triggers  & Compatible    &                   \\
\midrule
This work   & \checkmark    & \checkmark    & \checkmark    & \checkmark    & \checkmark    \\
(RADAR)     &               &               &               &               &               \\
\midrule
Weighted                 & $\times$     & \checkmark & \checkmark & $\times$ & \checkmark \\
Sampling~\cite{lascasas} & (clustering) &            &            & (custom) &  \\
\midrule
Diagnosis-           & $\times$   & $\times$        & \checkmark & $\times$ & \checkmark \\
Effective~\cite{xai} & (rules/ML) & (discards rare) &            & &            \\
\midrule
Blame-                    & $\times$ & $\times$ & $\times$          & $\times$                  & \checkmark \\
Proportional~\cite{audit} &          &          &  & (dynamic instr.) & \\
\midrule

eBPF-                         & $\times$ & $\times$ & \checkmark & \checkmark & \checkmark \\
Enhanced~\cite{ebpf-enhanced} &          &          &  & &  \\
\midrule

Hubble~\cite{hubble} & $\times$ & $\times$ & $\times$          & $\times$ & $\times$ \\
                     &          &          &  & (custom) & (Android) \\
\midrule

Hindsight~\cite{hindsight} & $\times$ & \checkmark & $\times$          & \checkmark & \checkmark \\
                           &          &            &  &  &  \\

\bottomrule
\end{tabular}
\end{table}

\section{Proposal}\label{proposal}

This section describes the methodological approach adopted for the design and implementation of \textbf{RADAR} (Reinforcement Learning Agent for Dynamic And Relevant trace sampling). The methodology rests on four pillars: (1) grounding in observability standards; (2) formalizing the sampling problem as an RL environment; (3) using Shannon entropy as the informational relevance metric; and (4) implementing an automated feedback loop on containerized infrastructure. The central goal is to let the system autonomously learn which sets of sampling rules preserve the data with the highest diagnostic utility, reducing observability cost.

\subsection{RADAR system architecture}

\begin{figure}
    \centering
    \includegraphics[width=0.85\linewidth]{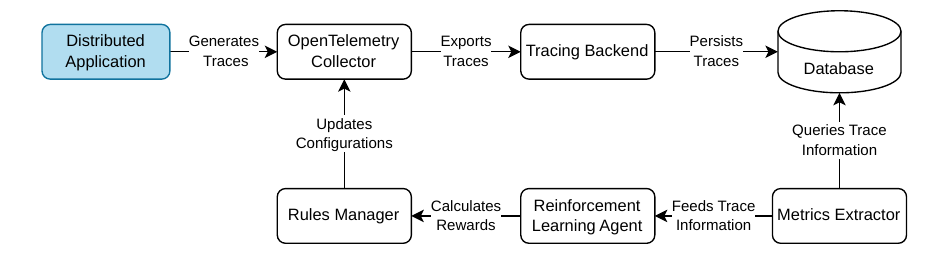}
    \caption{RADAR architecture: the Distributed Application's traces flow through the OpenTelemetry Collector to a Tracing Backend and Database; the Metrics Extractor queries this data to compute entropy and volume statistics for the Reinforcement Learning Agent, which selects a new sampling policy; the Rules Manager then applies this decision as an updated Collector configuration, closing the loop.}
    \label{fig:arquitetura_agente}
\end{figure}

The proposed architecture relies on a closed feedback loop connecting the monitored application to an autonomous decision-making pipeline, as illustrated in Figure~\ref{fig:arquitetura_agente}. The \textbf{Distributed Application} generates traces during normal operation, which are exported to an \textbf{OpenTelemetry Collector} enforcing the currently active sampling rules. The Collector exports the resulting traces to a \textbf{Tracing Backend}, which persists them in a \textbf{Database}. A \textbf{Metrics Extractor} queries this database to compute statistical features of the collected traces -- namely their entropy and volume -- and feeds this information to the \textbf{Reinforcement Learning Agent}. The agent uses these statistics to compute the reward (Eq.~\eqref{eq:reward}) and select a new subset of sampling policies from its catalog. Finally, a \textbf{Rules Manager} translates the agent's decision into an updated configuration for the OpenTelemetry Collector, closing the loop for the next episode. The concrete tools implementing the Tracing Backend, Database, and Rules Manager are described in Section~\ref{prototype}.

This architecture instantiates the classical MAPE-K autonomic control loop~\cite{kephart2003vision}: the OpenTelemetry Collector and Tracing Backend/Database jointly realize the \textit{Monitor} stage, continuously collecting traces from the Distributed Application; the Metrics Extractor performs the \textit{Analyze} stage, deriving entropy and volume statistics from the raw trace data; the Reinforcement Learning Agent realizes the \textit{Plan} stage, using the reward signal (Eq.~\eqref{eq:reward}) to select a new sampling configuration; and the Rules Manager performs the \textit{Execute} stage, applying this configuration back to the Collector. The agent's learned policy probabilities (Section~\ref{prototype}) act as the shared \textit{Knowledge} that persists and is refined across episodes.

\subsection{Reinforcement learning problem formalization}

As RADAR acts on the telemetry collector's global configuration without observing the environment's prior state (such as the exact CPU usage, memory, or traffic volume at the moment of decision), the adaptive sampling challenge does not constitute a traditional Markov Decision Process (MDP): there is no observable state on which the agent could condition its decisions, and therefore no state transition for it to learn. Instead, the problem is formalized as a \textbf{combinatorial Multi-Armed Bandit}, defined by the tuple $(A, R)$: an action space $A$ and a reward function $R$.

To solve this problem, we adapt the \textbf{REINFORCE} policy-gradient algorithm to a stateless setting, in which the agent learns the best probability distribution for selecting rule combinations based solely on the history of rewards obtained.

The action space $A$ is discrete and combinatorial. RADAR maintains a predefined catalog of $N$ independent sampling policies (e.g., rules based on an error status code, latency above a threshold, or business-specific attributes such as cart value). At each episode $t$, the agent's action $a_t \in \{0,1\}^N$ consists of selecting a binary vector, where each position $i$ indicates whether the corresponding policy is active ($a_{t,i}=1$) or inactive ($a_{t,i}=0$) in the OpenTelemetry Collector's configuration. Listing~\ref{lst:policy-example} shows an example of one such policy, expressed as the JSON rule the Collector consumes. The probability of selecting each rule is maintained internally by the agent and updated via gradient ascent.

\begin{lstlisting}[caption={Example sampling policy (a latency-based rule)}, label={lst:policy-example}]
{
  `name': `latency-policy-500ms',
  `type': `latency',
  `latency': { `threshold_ms': 500 }
}
\end{lstlisting}

The reward function $R$ is the core of this methodology, designed to balance information richness against operational cost. It is formalized as:
\begin{equation}
    R = \alpha \cdot H(X) - \beta \cdot P(n)
    \label{eq:reward}
\end{equation}
where $H(X)$ is the Shannon entropy of the collected traces (Eq.~\eqref{eq:shannon}), measuring the diversity of the information gathered -- higher entropy indicates the system is collecting varied, rare traces rather than repeating common, trivial paths. $P(n)$ is a volume penalty -- a sigmoid- or logistic-shaped function of the number of collected traces $n$ -- designed to keep the collected volume under an acceptable threshold $C$: the penalty stays low while $n \ll C$, but grows sharply as $n$ approaches $C$. Through gradient ascent, the agent increases the probability of selecting rule combinations that yield high entropy at low volume. $\alpha$ and $\beta$ are weighting hyperparameters that control the trade-off between information diversity and resource savings: increasing $\alpha$ adapts RADAR to environments more tolerant of cost, while increasing $\beta$ suits highly resource-constrained environments.

\subsection{Informational relevance via entropy}

To quantify data usefulness without human supervision, RADAR relies on \textbf{Shannon entropy} (Section~\ref{fundamentals}). In the context of telemetry, entropy measures the variability of execution paths. To compute it, traces -- complex trees of spans -- are converted into deterministic textual representations. This conversion involves hierarchically ordering the spans and filtering out irrelevant high-cardinality attributes (such as dynamic IP addresses); additionally, continuous values such as latency are discretized into intervals (buckets) to prevent natural network variation from artificially inflating the metric. The final entropy is computed over the frequency distribution of these unique trace representations, allowing the agent to prioritize anomalous behavior.

\subsection{Training loop}

RADAR's training proceeds as an iterative, episode-based process. At the start of each episode, the agent selects a new sampling configuration according to its current policy probabilities. This configuration is then deployed to the telemetry collector, and the system waits for a stabilization period before beginning measurement, ensuring the new configuration is fully active (the mechanics of this deployment step are detailed in Section~\ref{prototype}). Traces collected under the new policy during this window are converted into their textual representations and used to compute the frequency distribution -- and hence the entropy $H(X)$ -- of the observed data, alongside the collected volume $n$. The resulting reward $R$ is then used to update the agent's rule-selection probabilities via gradient ascent, increasing the likelihood of selecting combinations that yield high entropy at low volume in future episodes.

\section{Prototype}\label{prototype}

This section details the software implementation of the proposed system, translating the mathematical and architectural models described in Section~\ref{proposal} into functional components. The implementation is written in Python, using libraries for container orchestration (Kubernetes), numerical computation (NumPy), and interaction with the observability backend (Elasticsearch).

\subsection{Experimental infrastructure}

RADAR's software components are deployed alongside a standard distributed tracing stack, illustrated in the top panel of Figure~\ref{fig:arquitetura_boutique}. Traces generated by the monitored application are received by an \textbf{OpenTelemetry Collector}, which applies the currently active sampling rules and exports the retained traces to \textbf{Jaeger} -- the concrete Tracing Backend introduced in Section~\ref{proposal} -- for storage and visualization. Jaeger persists trace data in \textbf{Elasticsearch}, which serves as the Database queried by the Metrics Extractor (\texttt{es\_utils.py}). The RADAR modules (\texttt{agent.py}, \texttt{manager.py}, and \texttt{es\_utils.py}) close the loop by querying Elasticsearch, computing the reward, and updating the Collector's configuration through the Rules Manager, as detailed below. The full stack -- application, tracing backend, and RADAR modules -- can be deployed locally via Docker Compose or across a Kubernetes cluster; the experiments reported in Section~\ref{evaluation} used the Kubernetes deployment.

\subsection{RADAR implementation}

The codebase is organized into three main modules: the \textbf{RADAR agent} (\texttt{agent.py}), the \textbf{Rules Manager} (\texttt{manager.py}), and the \textbf{Metrics Extractor} (\texttt{es\_utils.py}).

The system's decision-making core is encapsulated in the \texttt{ReinforceAgent} class. Rather than a traditional Q-learning approach with tables over discrete state spaces, RADAR implements the \textbf{REINFORCE} policy-gradient algorithm, adapted to the combinatorial Multi-Armed Bandit formulation introduced in Section~\ref{proposal}, since the environment's ``state'' does not vary explicitly between iterations -- only the active policy configuration does.

The action space is modeled as a vector of independent probabilities, where each sampling policy in the catalog (defined in \texttt{tail\_sampling\_policies.json}) has a probability $p_i$ of being activated. In the \texttt{select\_actions} method, the decision to activate each policy is made through an independent Bernoulli sample for every item in the catalog:

\begin{lstlisting}[language=Python, caption={Action selection}, label={lst:action-sampling}]
if np.random.rand() < self.probs[i]:
    selected.append(policy)
    actions.append(1)
\end{lstlisting}

To encourage exploration and avoid premature convergence to local minima, probabilities are clipped to the $[0.01, 0.99]$ range during the update step, ensuring no policy is ever permanently disabled or permanently fixed.

The update rule applies gradient ascent to adjust these probabilities based on the received reward. A baseline, computed as an exponential moving average (EMA) of past rewards, is used to reduce the variance of the gradient estimate and stabilize learning:
\begin{equation}
    A_t = R_t - b_t, \qquad b_t = \gamma \cdot b_{t-1} + (1-\gamma) \cdot R_t
    \label{eq:advantage}
\end{equation}
where $A_t$ is the advantage at step $t$ -- positive if the chosen action performed better than the agent's recent historical average, negative otherwise -- $R_t$ is the reward returned for the current episode (Eq.~\eqref{eq:reward}), and $b_t$ is the EMA baseline, with $\gamma$ (\texttt{baseline\_decay}, which defaults to 0.9 in our implementation) controlling how much short-term reward fluctuations are smoothed. The probability vector is then updated in the direction of the estimated policy gradient:
\begin{equation}
    \theta_{t+1} = \theta_t + \eta \cdot A_t \cdot (a_t - \theta_t)
    \label{eq:policy-update}
\end{equation}
where $\theta_t$ is the vector of per-policy probabilities (\texttt{self.probs}), $\eta$ is the learning rate (\texttt{lr}, which defaults to 0.2 in our implementation), and $(a_t - \theta_t)$ approximates the gradient direction: if policy $i$ was activated ($a_{t,i}=1$) under a positive advantage, this term pushes $\theta_{t,i}$ toward 1; if it was not activated ($a_{t,i}=0$) under a positive advantage, its probability is reduced instead.

\subsection{Reward function implementation}

The \texttt{reward} function implements the multi-objective logic introduced in Section~\ref{proposal}, combining a normalized entropy term with a sigmoid-shaped volume penalty:

\begin{lstlisting}[language=Python, caption={Reward function implementation}, label={lst:reward-function}]
def trace_penalty(traces, C, k=25, midpoint=0.20):
    x = traces / C
    return 1 / (1 + math.exp(-k * (x - midpoint)))

def reward(entropy, traces, alpha=1.0, beta=1.0, C=10000):
    norm_entropy = entropy / 10
    penalty = trace_penalty(traces, C)
    return alpha * norm_entropy - beta * penalty
\end{lstlisting}

\begin{figure}[h]
    \centering
    \includegraphics[width=0.8\textwidth]{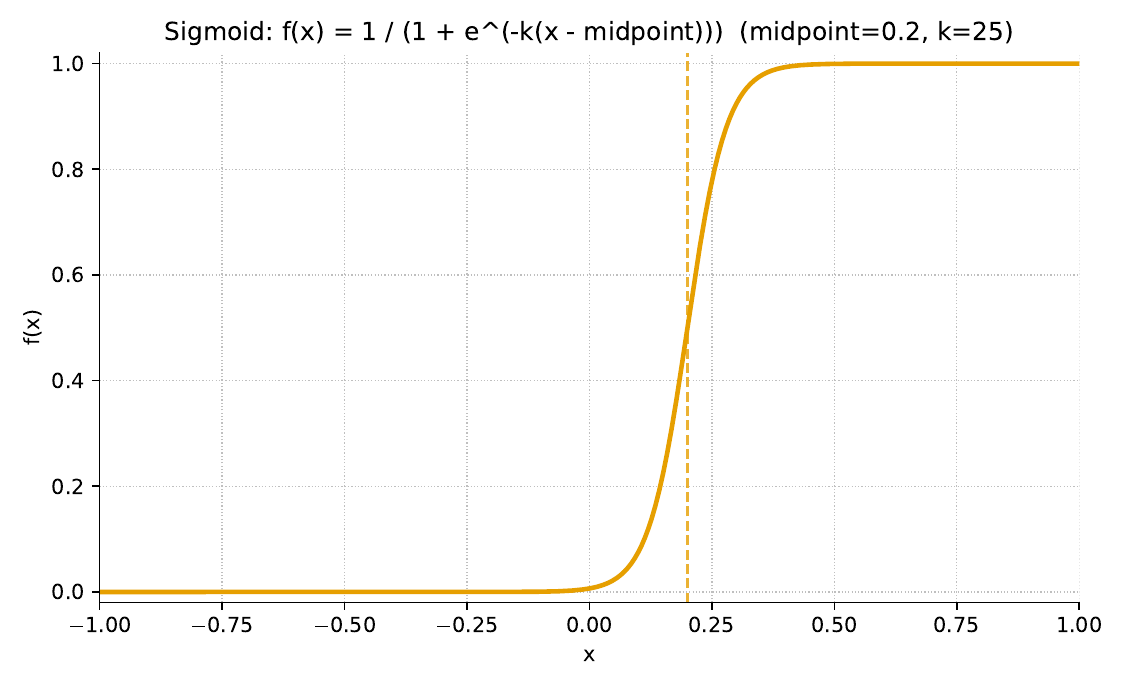}
    \caption{Sigmoid penalty function for trace volume.}
    \label{fig:sigmoid}
\end{figure}

Here, \texttt{alpha} and \texttt{beta} are the $\alpha$ and $\beta$ weighting hyperparameters of Eq.~\eqref{eq:reward}; in this implementation they default to $\alpha=1.0$ and $\beta=1.0$, with the volume threshold set at $C=10{,}000$ traces. The penalty function is a logistic curve $1/(1+e^{-k(x-\text{midpoint})})$, shown in Figure~\ref{fig:sigmoid} for $k=25$ and $\text{midpoint}=0.20$: it penalizes trace volume mildly below the threshold and sharply beyond it, letting RADAR operate freely under the cost budget while suffering a strong penalty for exceeding it.

\subsection{Metrics extraction and entropy calculation}

Data quality evaluation is performed by the \texttt{es\_utils.py} module, which queries Elasticsearch to retrieve the traces tagged with the current experiment's hash. Computing Shannon entropy requires converting each trace's complex, hierarchical span tree into a comparable, discrete (string) representation. This is done by the \texttt{trace\_to\_string} function in three steps: (1) \textbf{ordering} -- spans are sorted hierarchically (parents before children) and temporally, so the same logical execution always yields the same string; (2) \textbf{noise filtering} -- high-cardinality attributes irrelevant to behavioral structure (e.g., \texttt{span.kind}, \texttt{peer\_port}, dynamic IP addresses) are removed via a configurable \texttt{tag\_blacklist}; and (3) \textbf{quantization} -- continuous values such as latency (\texttt{duration\_ms}) are discretized into buckets (e.g., 200 ms intervals) by \texttt{quantize\_value\_if\_applicable}, preventing small, natural network timing variations from being interpreted as distinct behaviors and artificially inflating entropy.

Once traces are converted into unique strings, their frequency distribution is used to compute entropy:

\begin{lstlisting}[language=Python, caption={Entropy calculation}, label={lst:entropy-calculation}]
def calculate_entropy(traces):
    strings = []
    for trace_id, spans in traces.items():
        s = trace_to_string(spans)
        strings.append(s)

    if not strings:
        return 0.0

    counter = Counter(strings)
    total = sum(counter.values())
    ps = [c / total for c in counter.values()]

    alpha = ENTROPY_ALPHA

    if abs(alpha - 1.0) < 1e-12:
        entropy = -sum(p * math.log2(p) for p in ps if p > 0)
        return entropy

    sum_p_alpha = sum((p ** alpha) for p in ps)
    sum_p_alpha = max(sum_p_alpha, 1e-300)
    entropy = (1.0 / (1.0 - alpha)) * math.log2(sum_p_alpha)
    return entropy
\end{lstlisting}

Note that this implementation is slightly more general than the Shannon entropy defined in Section~\ref{fundamentals}: it computes the Rényi entropy of order \texttt{ENTROPY\_ALPHA}, which reduces exactly to Shannon entropy when \texttt{ENTROPY\_ALPHA}$=1$ -- the configuration used throughout this work.

\subsection{Deployment orchestration}

The \texttt{manager.py} script (the Rules Manager introduced in Section~\ref{proposal}) acts as the main controller, implementing the experiment's lifecycle as a continuous loop. It is responsible for interfacing with the Kubernetes API and applying RADAR's decisions to the live environment.

At the start of each episode, the current collector configuration file must be replaced. The \texttt{generate\_config} function translates RADAR's abstract decision into a manifest the OpenTelemetry Collector understands: it builds a Python dictionary mirroring the collector's required hierarchy (\texttt{receivers}, \texttt{processors}, \texttt{exporters}, \texttt{service} sections), injects the list of policies selected by RADAR directly into \texttt{processors} $\rightarrow$ \texttt{tail\_sampling} $\rightarrow$ \texttt{policies}, and tags the configuration with a hash of the current experiment in \texttt{processors} $\rightarrow$ \texttt{attributes} -- so every trace the collector processes is automatically annotated with an \texttt{experiment\_hash}, letting the Metrics Extractor later identify which rule set produced it. The dictionary is then serialized to YAML and written to the environment's Kubernetes \texttt{ConfigMap}.

Simply updating the ConfigMap is not enough, since running pods do not automatically reload static configuration. To force an update without downtime, the system uses the Kubernetes API to patch the Collector's \texttt{Deployment} object, inserting the new configuration hash as a \texttt{config-hash} annotation in \texttt{spec.template.metadata.annotations}. Kubernetes interprets any change to the Pod template specification -- even a metadata-only annotation -- as a change to the application definition, and automatically triggers a rolling update: a new \texttt{ReplicaSet} is created with the updated template, new pods are started gradually (mounting the already-updated ConfigMap as they come up), and old pods are terminated only once the new ones are ready. \texttt{manager.py} then enters a polling loop against the Kubernetes API, waiting until the number of available replicas matches the desired count, ensuring RADAR only begins measuring the environment once the new configuration is fully active. Table~\ref{tab:fluxo-radar} summarizes this episode lifecycle. The RADAR implementation is publicly available at \url{https://github.com/ComputerNetworks-UFRGS/RADAR}.

\begin{table}[h]
\centering
\caption{Implementation flow: RADAR's episode lifecycle}
\label{tab:fluxo-radar}
\begin{tabular}{llp{6.5cm}}
\toprule
\textbf{Step} & \textbf{Component} & \textbf{Action} \\
\midrule
1. Initialization & Python script & Generates the dynamic YAML file + hash. \\
2. Persistence & K8s ConfigMap & Updates the ConfigMap with the new OTel policy. \\
3. Trigger & K8s API (patch) & Updates the Deployment, injecting the hash into its metadata. \\
4. Orchestration & K8s Controller & Detects the change and starts the Rolling Update. \\
5. Stabilization & New OTel Pods & The Collector comes up and starts measuring once ready. \\
\bottomrule
\end{tabular}
\end{table}

\section{Experimental Evaluation}\label{evaluation}

This section presents the experimental evaluation of RADAR, covering the test environment, the agent's convergence behavior, resource consumption, and its effectiveness at preserving rare, diagnostically valuable traces.

\subsection{Experimental environment}

\begin{figure}[ht]
    \centering
    \includegraphics[width=0.85\linewidth]{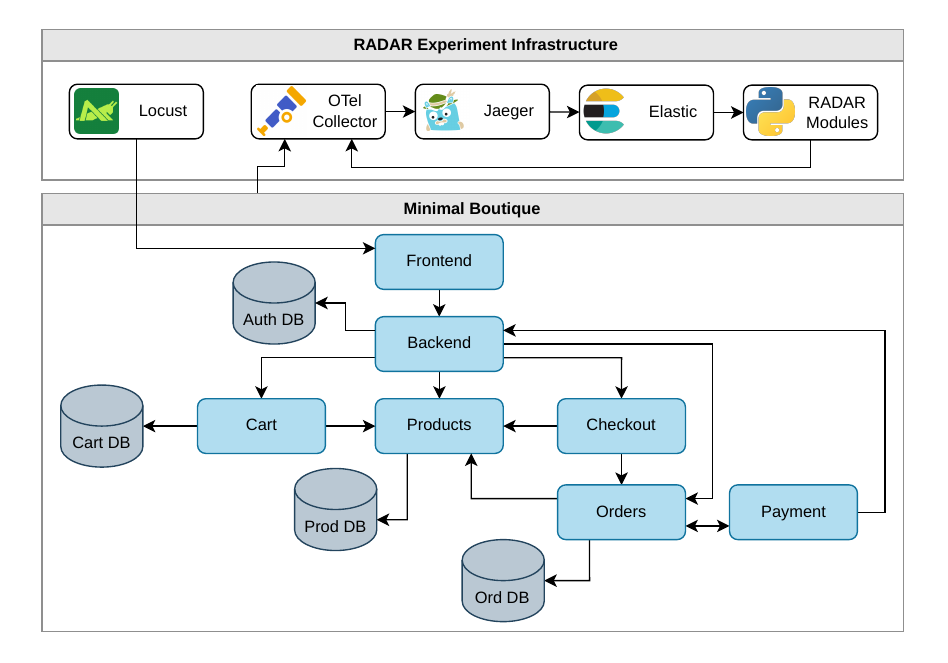}
    \caption{RADAR experiment infrastructure and Minimal Boutique architecture. Top: Locust generates load against the application; traces flow from the OpenTelemetry Collector to Jaeger and Elasticsearch, from which the RADAR modules compute the reward and update the Collector's configuration, closing the loop. Bottom: Minimal Boutique's service topology, with the Frontend routing through the Backend (API gateway/authentication) to the Cart, Products, Checkout, Orders, and Payment services, each backed by its own database.}
    \label{fig:arquitetura_boutique}
\end{figure}

RADAR was validated on \textbf{Minimal Boutique}, a microservices-based online store distributed across a Kubernetes cluster.\footnote{Minimal Boutique is available at \url{https://github.com/ComputerNetworks-UFRGS/minimal-boutique}.} Minimal Boutique was developed in-house by this research group specifically to retain full control over the source code and service topology, allowing an isolated baseline environment in which anomalies detected by RADAR could be attributed exclusively to the scenarios produced by the experiment, without noise intrinsic to third-party benchmarks. It is a reduced, optimized adaptation of the \textit{Google Online Boutique\footnote{Google Online Boutique is available at: \url{https://github.com/googlecloudplatform/microservices-demo}.}} reference application, focused specifically on generating rich telemetry (distributed traces) for observability experiments. The application follows the \textit{Database-per-Service} pattern, forcing all inter-service communication through HTTP/REST APIs and maximizing the generation of network spans, essential for testing the capture of distributed latencies. It comprises seven services -- a web \textbf{frontend}; a \textbf{backend} acting as API gateway and authentication layer; and the \textbf{products}, \textbf{cart}, \textbf{checkout}, \textbf{payment}, and \textbf{orders} services -- each implemented in Python with Flask and SQLAlchemy, sharing a common OpenTelemetry instrumentation module so that generated telemetry is structurally homogeneous across services. The payment service in particular simulates an external payment gateway and interacts with both the orders and cart/backend services to confirm and finalize purchases, introducing extra network hops that enrich the generated traces. The bottom panel of Figure~\ref{fig:arquitetura_boutique} shows the application's service topology, including each service's dedicated database, consistent with the Database-per-Service pattern.

To simulate realistic traffic, \textbf{Locust}\footnote{Locust: an open-source load testing tool, \url{https://locust.io/}} is configured to generate stochastic load. Virtual user behavior is modeled as a Markov chain, transitioning between browsing, adding items to the cart, and checkout with different probabilities, weighted to reflect a typical sales funnel (browsing most frequent, checkout least). This ensures a heterogeneous mix of telemetry: a large number of simple, repetitive traces and a small number of long, complex ones, challenging the agent's ability to select intelligently.

\subsection{Convergence experiments}

To evaluate how the method's efficiency scales with the number of rules available to the agent, four incremental experiments were conducted. Available rules were grouped into three categories: \textbf{structural rules}, related to overall service behavior and performance (e.g., latency above 500 ms, traces with more than 20 spans, database queries slower than 500 ms, all system errors); \textbf{probabilistic rules}, applied to fixed proportions of requests (e.g., 10\% or 20\% of all traces, 1\% of all state-changing requests, 1\% of all operations on the products service); and \textbf{business-specific rules}, configured for scenarios particularly relevant to the application domain (e.g., purchases above R\$500, declined payments, carts with more than 10 items, orders with more than 4 items). Table~\ref{tab:regras_incrementais} shows the four experiments, each incrementally adding rules from these categories.

\begin{table}[ht]
\centering
\caption{Rules selected per experiment (incremental)}
\label{tab:regras_incrementais}
\begin{tabular}{cp{8.5cm}c}
\toprule
\textbf{Experiment} & \textbf{Selected rules} & \textbf{Total} \\ \midrule
1 & Latency $>$ 500 ms; 10\% probability; purchases $>$ R\$500 & 3 \\ \hline
2 & Previous experiment's rules + traces with more than 20 spans; 20\% probability; declined payments & 6 \\ \hline
3 & Previous experiment's rules + slow database queries; 1\% of all system-changing operations; more than 10 items in cart & 9 \\ \hline
4 & Previous experiment's rules + errors; 1\% of all operations on the products service; more than 4 orders placed & 12 \\ \bottomrule
\end{tabular}
\end{table}

The following metrics were monitored to assess convergence: \textbf{cumulative reward}, indicating RADAR's performance at optimizing its collection criteria; \textbf{trace entropy}, representing the informational diversity that is the work's central objective; and the \textbf{total number of collected traces}, assessing the impact on observability cost.

\begin{figure}[ht]
    \centering
    \includegraphics[width=0.8\linewidth]{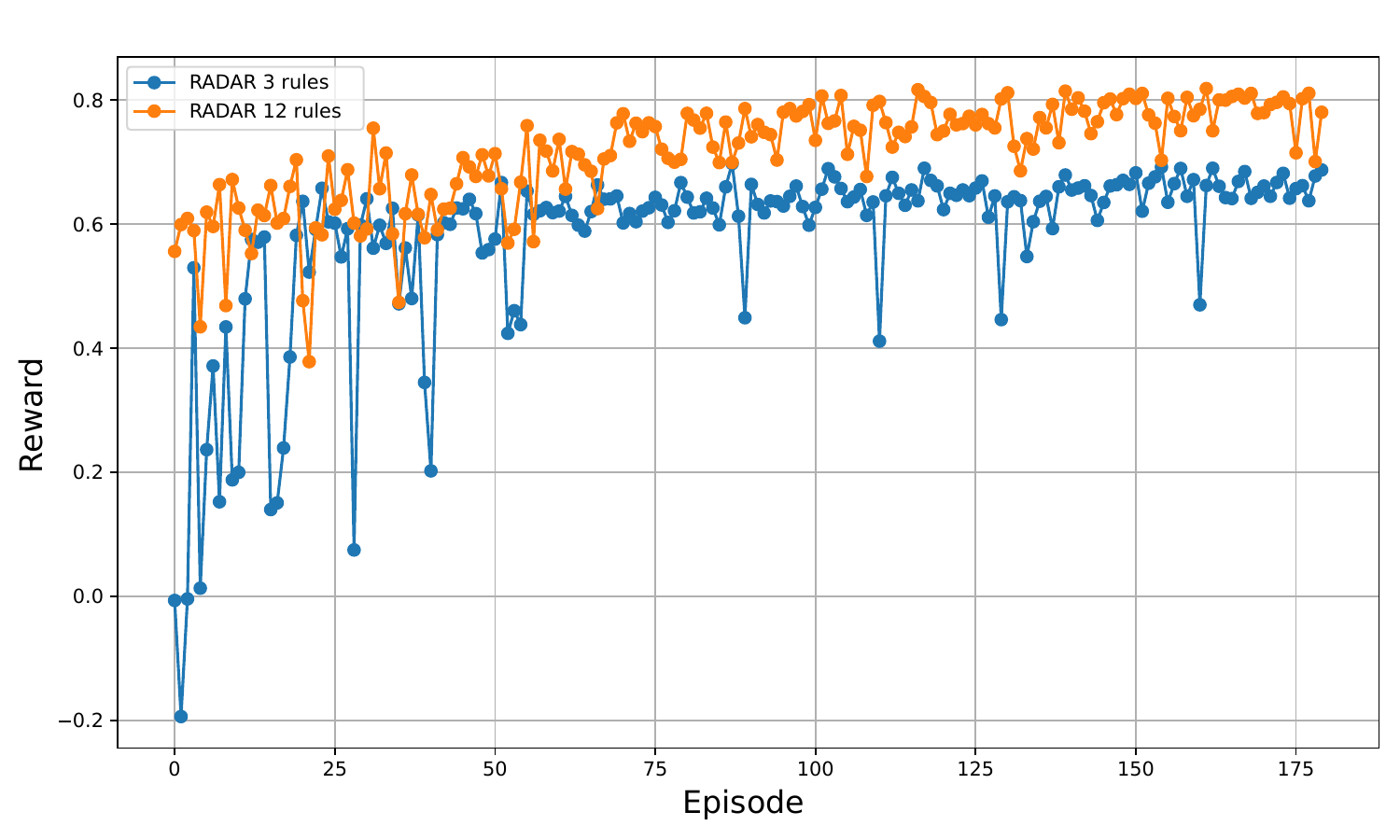}
    \caption{Reward convergence with 3 and 12 rules.}
    \label{fig:grafico_reward_5_12_3}
\end{figure}

\begin{figure}[ht]
    \centering
    \includegraphics[width=0.8\linewidth]{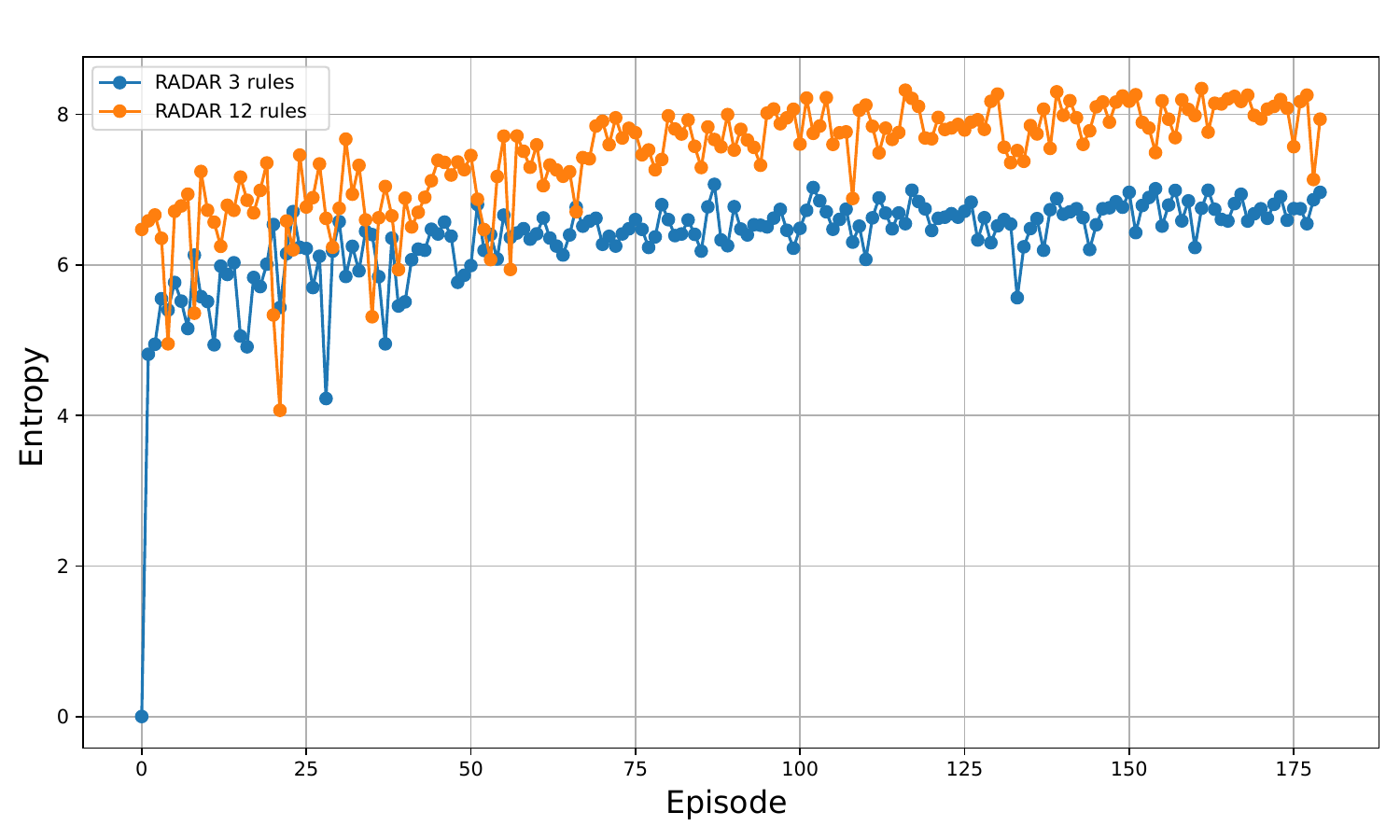}
    \caption{Entropy convergence with 3 and 12 rules.}
    \label{fig:grafico_entropia_5_3_12}
\end{figure}

\begin{figure}[ht]
    \centering
    \includegraphics[width=0.8\linewidth]{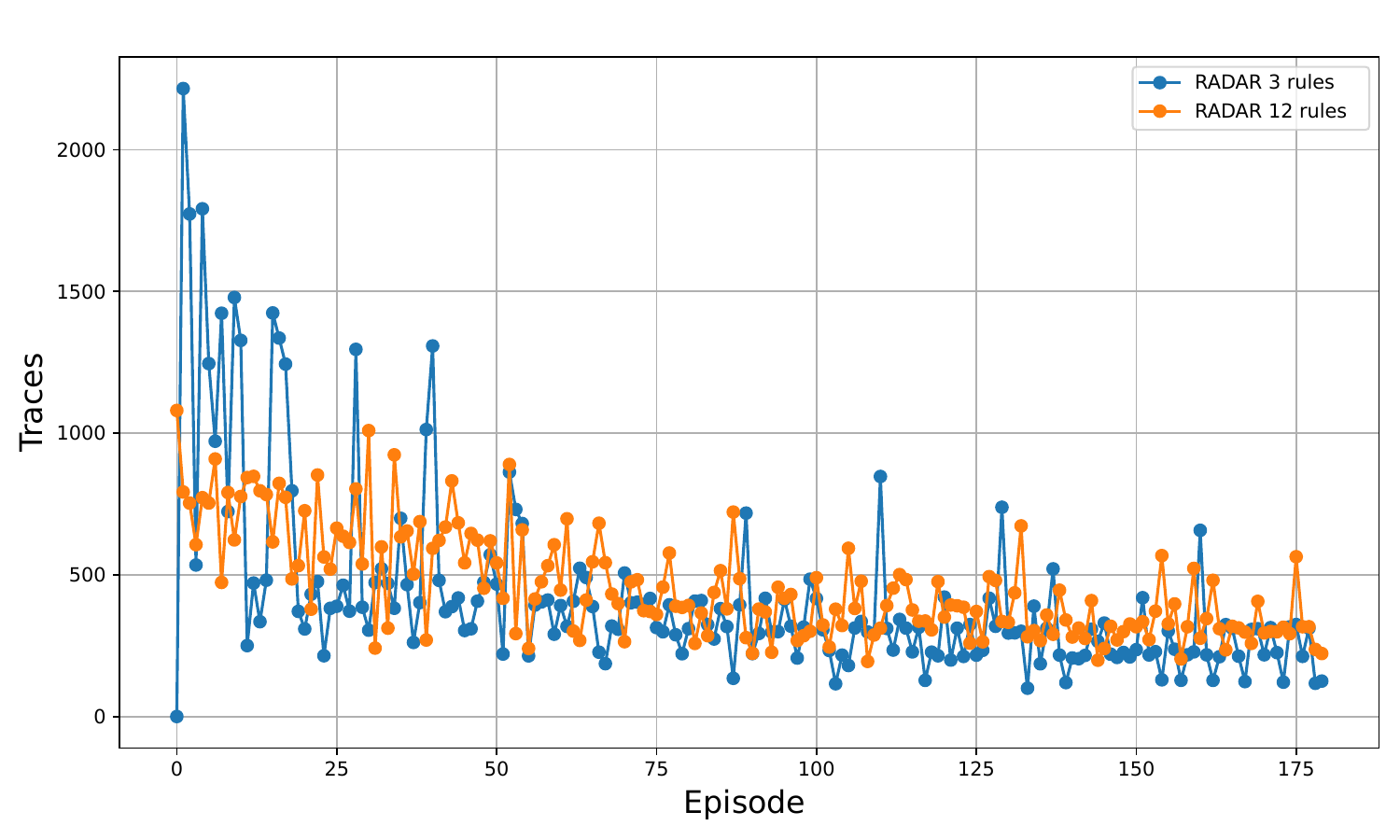}
    \caption{Number of traces collected per episode with 3 and 12 rules.}
    \label{fig:grafico_traces_5_3_12}
\end{figure}

Results show that, regardless of the number of rules available, RADAR converged to a stable policy, maximizing the reward received over episodes (Figure~\ref{fig:grafico_reward_5_12_3}); entropy was likewise maximized, converging to a value higher than at the start of the experiment (Figure~\ref{fig:grafico_entropia_5_3_12}). The experiment with three rules constrained the number of collected traces more sharply, having fewer options to explore, while the twelve-rule scenario reached higher levels of data richness at the cost of a longer convergence time (Figure~\ref{fig:grafico_traces_5_3_12}) -- interestingly, the twelve-rule configuration ultimately collected \textit{more} traces than the three-rule one, a cost the reward function deemed worth paying in exchange for higher entropy in the sampled data. For clarity, only the two most extreme cases (3 and 12 rules) are shown; the 6- and 9-rule scenarios were also tested and exhibited the same trends.


\subsection{Resource consumption}

To assess resource usage, three collector configurations were run for five hours each: a collector storing 100\% of received traces; a collector storing a fixed 20\% of received traces, which is the \textit{midpoint} value used in the volume penalty function (Eq.~\eqref{eq:reward}), and the maximum RADAR could reach before incurring heavy penalties; and RADAR itself, selecting policies via the trained agent.

\begin{figure}[ht]
    \centering
    \includegraphics[width=0.8\linewidth]{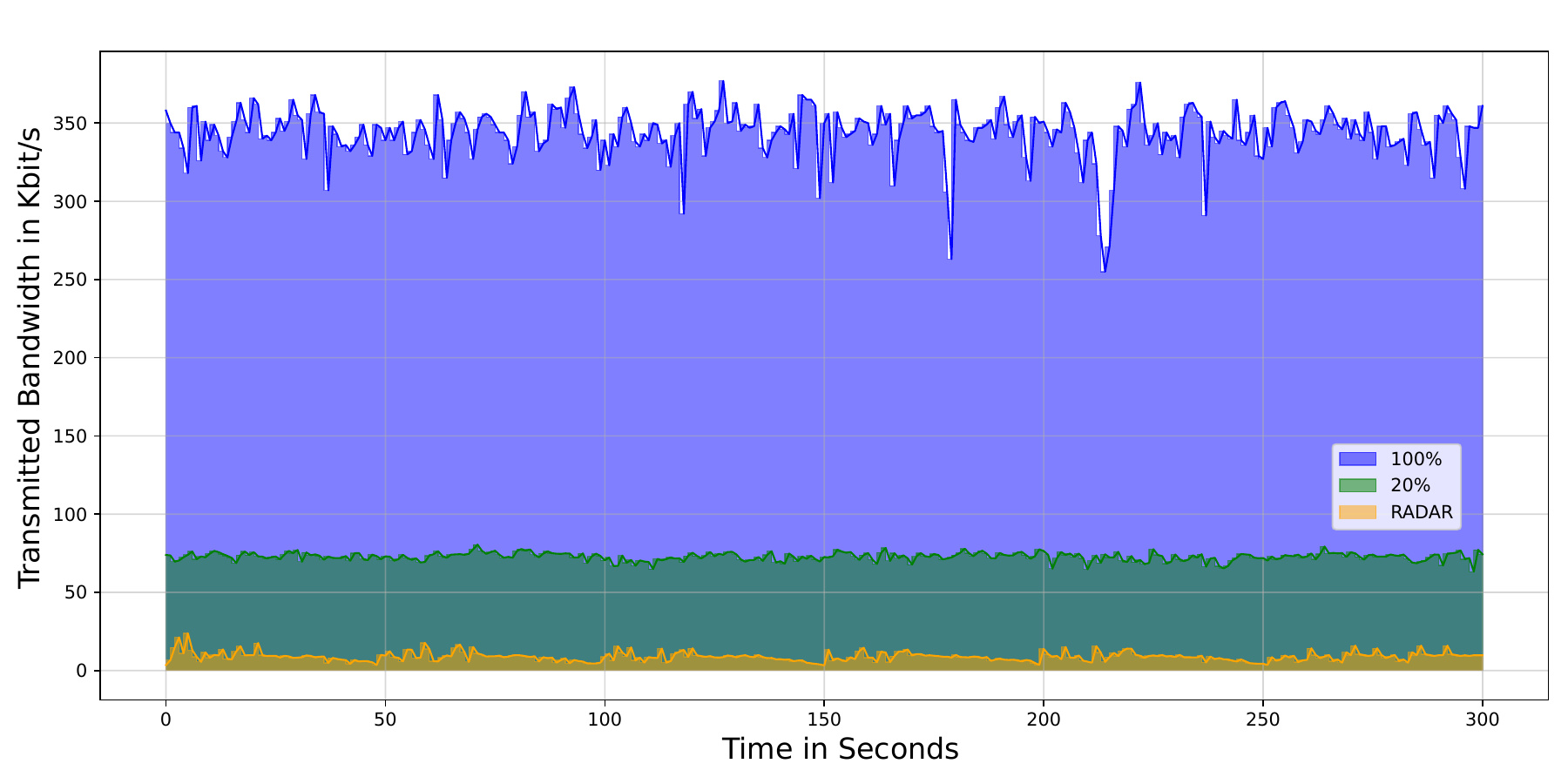}
    \caption{Bandwidth usage at 100\%, 20\%, and with RADAR.}
    \label{fig:desempenho_bandwidht}
\end{figure}

Figure~\ref{fig:desempenho_bandwidht} shows the average outbound bandwidth used by each collector over time. RADAR reduced bandwidth to an average of 9 Kbit/s, compared to 344 Kbit/s for the 100\% collector (a 97.4\% reduction) and 73 Kbit/s for the 20\% collector (an 87.7\% reduction), which is equivalent to a fixed sampling rule of roughly 2.5\%, without the probabilistic error-collection component.

\begin{figure}[ht]
    \centering
    \includegraphics[width=0.8\linewidth]{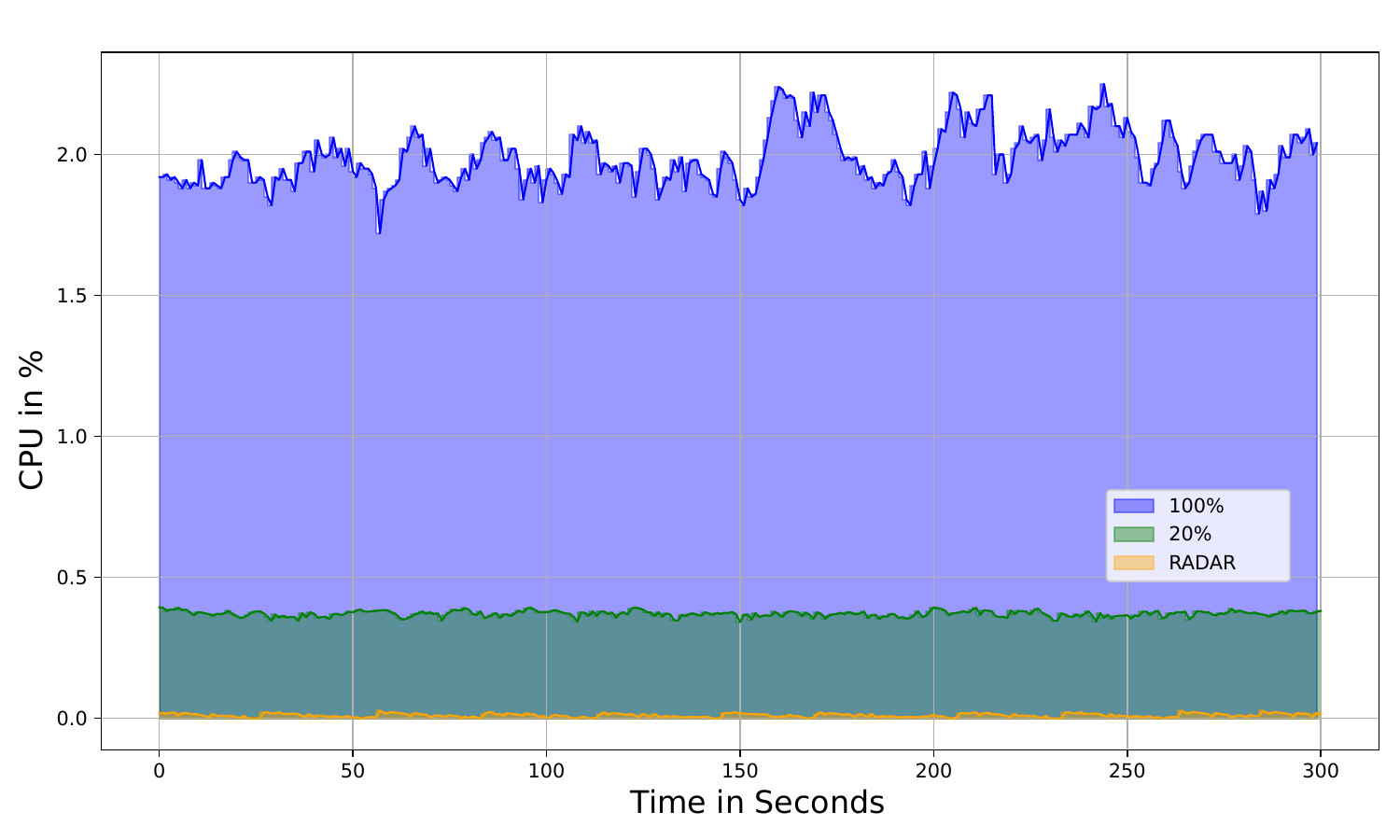}
    \caption{CPU usage at 100\%, 20\%, and with RADAR.}
    \label{fig:desempenho_cpu}
\end{figure}

Figure~\ref{fig:desempenho_cpu} shows the average CPU usage in percentage for each collector. The collector configured to store 100\% of traces used roughly an average of 1.99\% versus 0.02\% for RADAR, which represents a 99.0\% reduction in CPU usage. In turn, 0.38\% was the average CPU usage for the 20\% collector, against which RADAR still represents a 94.7\% reduction. These significant reductions are explained by the discarding of traces that are never processed and forwarded to Jaeger.

\begin{figure}[ht]
    \centering
    \includegraphics[width=0.8\linewidth]{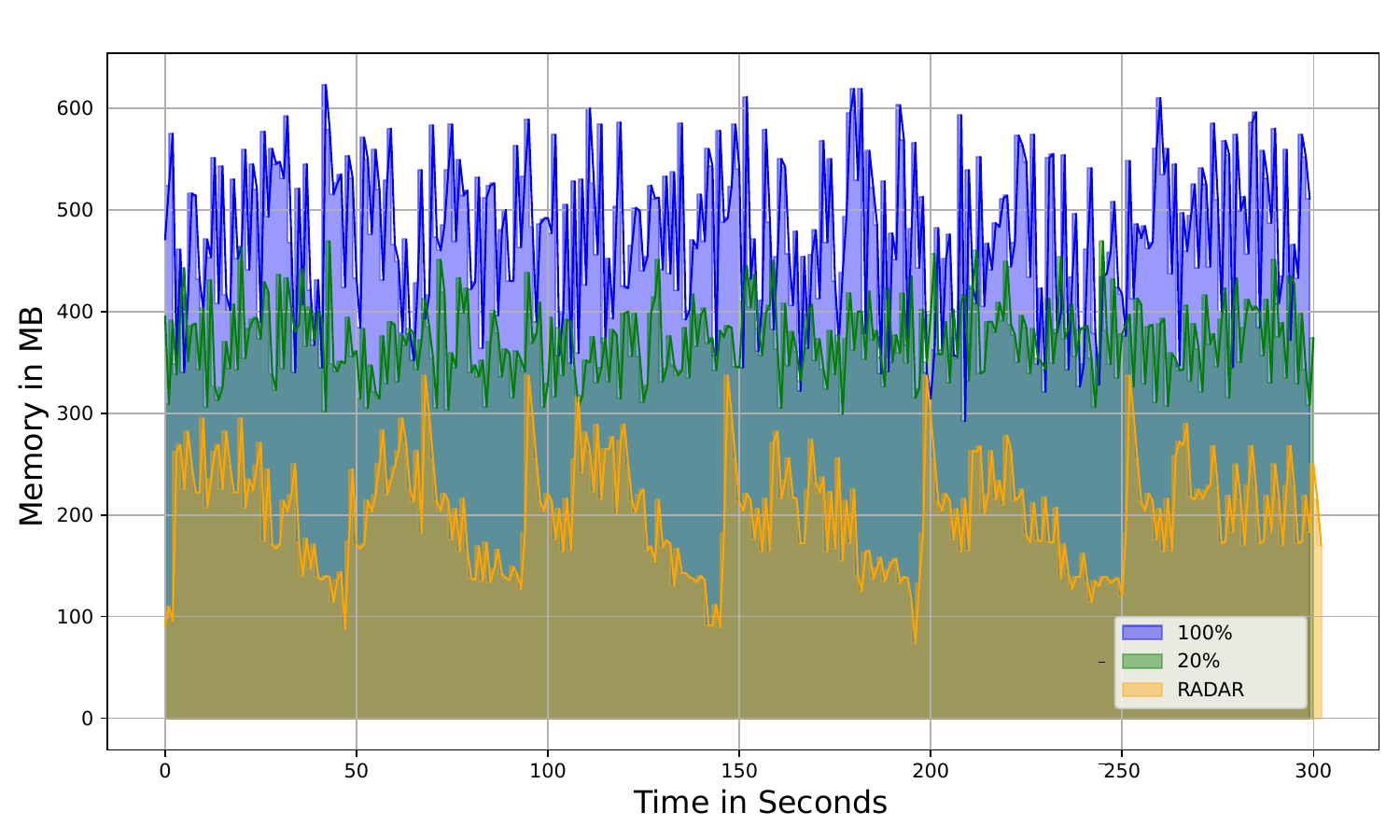}
    \caption{Memory usage at 100\%, 20\%, and with RADAR.}
    \label{fig:desempenho_mem}
\end{figure}

Figure~\ref{fig:desempenho_mem} shows average memory usage for the three collector configurations. The 100\% collector used 476 MB on average, while RADAR consumed roughly 200 MB (representing a 58.0\% reduction). Against the 20\% collector, which shows an average memory usage of 372 MB, RADAR uses 46.2\% less memory. 

\begin{figure}[ht]
    \centering
    \includegraphics[width=0.8\linewidth]{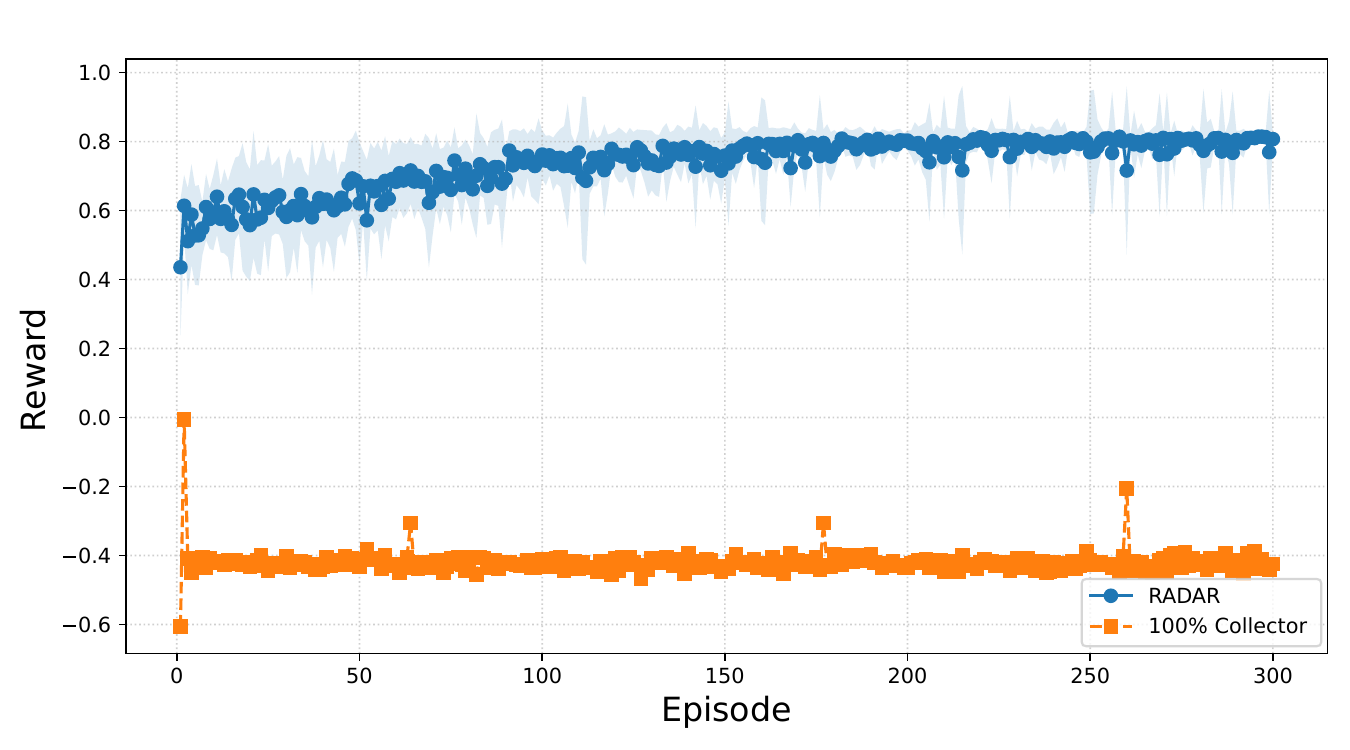}
    \caption{Reward evolution: RADAR vs. the 100\% collector.}
    \label{fig:g_100_vs_rl}
\end{figure}

\begin{figure}[ht]
    \centering
    \includegraphics[width=0.8\linewidth]{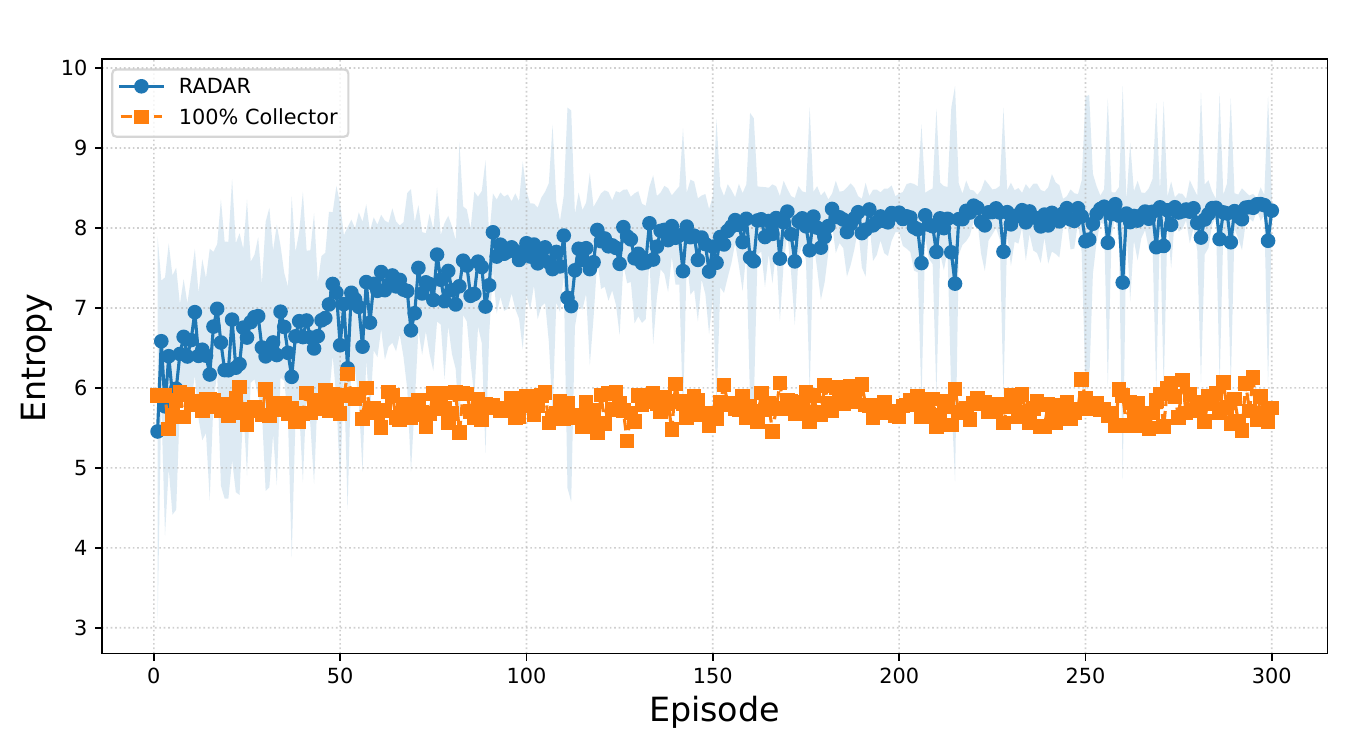}
    \caption{Entropy comparison: RADAR vs. the 100\% collector.}
    \label{fig:g_100_vs_rl_ent}
\end{figure}

\begin{figure}[ht]
    \centering
    \includegraphics[width=0.8\linewidth]{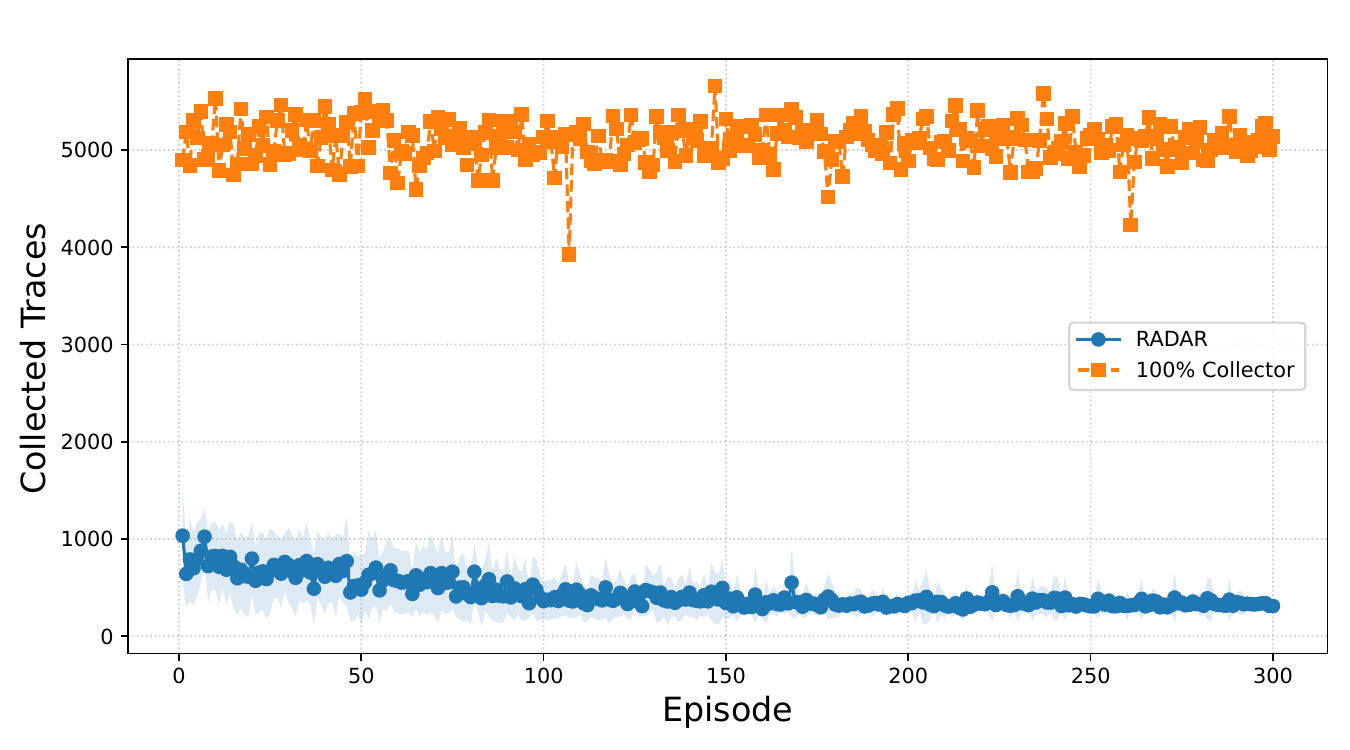}
    \caption{Traces collected per episode: RADAR vs. the 100\% collector.}
    \label{fig:g_100_vs_rl_trc}
\end{figure}

Figures~\ref{fig:g_100_vs_rl}--\ref{fig:g_100_vs_rl_trc} compare RADAR directly against the 100\% collector, episode by episode. The 100\% collector's reward remained stable around an average of $-0.43$, while RADAR's reward rose to an average of $0.74$, peaking at $0.79$ (Figure~\ref{fig:g_100_vs_rl}). Average entropy rose from 5.76 (100\% collector) to 7.63 with RADAR, an increase of approximately 24.5\% (Figure~\ref{fig:g_100_vs_rl_ent}). The average number of traces collected per episode fell from 5,083 (100\% collector) to 439 with RADAR, a reduction of approximately 91.4\% (Figure~\ref{fig:g_100_vs_rl_trc}).


\subsection{Rare trace preservation}

To test whether rare traces were effectively captured, an OpenTelemetry collector was configured with two independent processing pipelines: one using RADAR's selected rule set as the optimal configuration, and another configured to collect 100\% of traces. Each pipeline's output was tagged with an identifying key, \textit{baseline} for the full collection, \textit{experiment} for RADAR's configuration, to distinguish the two data flows during analysis.

After the collection period, traces from both pipelines were passed through RADAR's classification and normalization functions, converting them into canonical textual representations and removing unique variable attributes to reduce noise and enable structural comparison. From this classification and normalization, we can distinguish how many unique trace patterns are actually contained within the whole dataset (traces with the same canonical representation are considered redundant). In addition, the least frequent trace patterns were then ranked from the \textit{baseline} pipeline's dataset, and their presence was checked in the \textit{experiment} pipeline's collected set to compute a coverage percentage. We consider a trace to be rare if it appears only once during the entire test run.

The experiment was run across ten one-hour test cases with the dual pipeline. Table~\ref{tab:resumo-traces} shows the average traces collected and how many of those represent unique patterns.

\begin{table}[ht]
  \centering
  \caption{Trace summary}
  \label{tab:resumo-traces}
  \begin{tabular}{l r r}
  \toprule
    Type & \# traces & \# unique patterns \\ \midrule
    baseline & 484,967 & 63,237 \\
    experiment & 61,122 & 57,460 \\
    \bottomrule
  \end{tabular}
\end{table}

Unique patterns represent approximately 13.0\% of the baseline pipeline's total traces, compared to 94.0\% for RADAR's -- a direct consequence of RADAR discarding redundant, repetitive traces. Of the rare patterns (i.e., traces that appeared only once in a given test run) identified from the baseline pipeline, we looked into RADAR's collected traces to verify if they were also included there. We found that on average 85.6\% of rare traces were also present in RADAR's collected set, a high retention rate given the substantial reduction in both trace volume and the number of unique patterns collected.

\subsection{Discussion}

The results confirm that observability in microservices architectures presents a fundamental dilemma: the need to collect detailed data versus the prohibitive cost of processing the resulting volume of telemetry. RADAR's approach offers a viable strategy for balancing this dilemma by using entropy as a universal signal of interest.

\textbf{Efficiency and resource balance.} The 97.4\% reduction in bandwidth and the 99.0\% reduction in CPU usage show that intelligently discarding redundant data is more effective than collecting at a fixed probabilistic rate. While the 100\% collector generates a massive volume of repetitive data reflecting successful operations, RADAR learned to identify and prioritize the execution paths that genuinely add informational value. Compared to a fixed 20\% sampling rate, RADAR not only consumed fewer resources but did so while maximizing its reward, indicating that these savings did not come at the cost of randomly discarding information, but rather through qualitative filtering; demonstrating the viability of a multi-objective reward function that penalizes excessive volume while rewarding structural diversity in the collected traces.

\textbf{Observability preservation and informational value.} Retaining 85.6\% of rare trace patterns is one of the most significant results, since rare events are precisely the ones most relevant for diagnosis and anomaly detection. By increasing the average entropy of stored information by approximately 24.5\%, RADAR demonstrated its ability to distinguish between repetitive, redundant traces and anomalous ones. Unlike systems such as Hindsight, which operate reactively and depend on predefined anomaly triggers, RADAR acts proactively: the agent discovered which traces were relevant through entropy maximization alone, capturing scenarios of interest that would not have activated a conventional static trigger. Moreover, by operating entirely in user space through OpenTelemetry, RADAR ensures greater portability and avoids the complexity of dynamic code injection or invasive kernel modifications.

\textbf{Viability of autonomous orchestration.} The stable convergence of the policy, even as the number of rules grows, reinforces RADAR's potential as a practical, loosely coupled solution for modern distributed systems. The methodology showed that it is possible to drastically reduce data volume without sacrificing the observability of critical scenarios, validating the feasibility of using entropy to orchestrate telemetry autonomously and efficiently. We conclude that entropy-based intelligent agents offer a more dynamic balance than the manual tail-sampling configurations that are difficult to maintain in dynamic cloud environments.

\section{Conclusion}\label{conclusion}

Observability in microservices architectures presents a fundamental dilemma: the need to collect detailed data for failure diagnosis versus the prohibitive cost of storing and processing the resulting volume of telemetry. Traditional approaches, such as probabilistic (head-based) sampling, prove ineffective by indiscriminately discarding rare and critical events, while manually configured tail-sampling rules are difficult to maintain in dynamic environments.

This work addressed this challenge through RADAR (Reinforcement Learning Agent for Dynamic And Relevant trace sampling), a framework that combines reinforcement learning with information theory to dynamically adjust sampling policies. By using Shannon entropy as a diversity metric, the agent distinguishes between repetitive, redundant traces and information-rich, anomalous ones, adjusting the OpenTelemetry collector's rules autonomously and without requiring direct user intervention.

Experimental results confirmed the efficacy of this approach: RADAR reduced bandwidth consumption by 97.4\% and CPU usage by 99.0\% compared to full data collection, also outperforming a fixed 20\% sampling rate in resource efficiency. Informational quality was preserved: the system retained approximately 85.6\% of rare trace patterns while increasing the average entropy of stored data by approximately 25\%. These results validate the hypothesis that entropy-based autonomous orchestration offers a superior dynamic balance compared to conventional strategies, sustaining high observability at a drastically reduced computational cost.

We conclude that the application of entropy-based intelligent agents is a viable and efficient strategy for telemetry orchestration in distributed systems, offering a dynamic balance between operational cost and informational value.

\subsection{Contributions}

The main contributions of this work can be summarized as follows:
\begin{itemize}
    \item \textbf{Autonomous sampling mechanism (RADAR)}: an agent capable of interacting directly with the OpenTelemetry collector, eliminating the need for manual, static configuration of tail-sampling rules.
    \item \textbf{Entropy-based reward function}: a reward formulation that penalizes excessive volume while rewarding structural diversity in traces, proving to be an effective method for qualifying telemetry data ``usefulness'' without human supervision.
    \item \textbf{Experimental testbed}: the creation and public availability of an instrumented, containerized microservices testbed capable of generating realistic workloads, serving as a basis for future observability experiments.
\end{itemize}

\subsection{Threats to validity}

Despite these promising results in resource reduction and entropy preservation, this study has several limitations. First, regarding external validity, the experimental evaluation was conducted exclusively on Minimal Boutique, an environment built by the research group itself; although the load generator simulates a realistic e-commerce environment, the architecture and communication patterns reflect a single application domain, so the agent's convergence stability and optimal hyperparameter configuration may vary under more complex topologies or unforeseen workloads. Second, the evaluation does not include a comparison against a static configuration using the same rule catalog with fixed or uniform activation probabilities. Such a comparison could help more directly isolate the contribution of the learned policy from the value of the underlying rule catalog itself. Third, the reward function's hyperparameters ($\alpha$, $\beta$, $C$, $k$, and $\text{midpoint}$, Section~\ref{prototype}) were selected through manual experimentation rather than systematic sensitivity analysis. While the reported configuration produced stable, reproducible behavior throughout our experiments, we do not characterize how performance degrades outside this configuration. Finally, each policy update triggers a Kubernetes rolling update of the collector (Section~\ref{prototype}), and this work does not quantify the operational cost or transient disruption of that reconfiguration step itself, which would be relevant for assessing RADAR's overhead in a production setting. Generalizing RADAR's empirical model to heterogeneous production systems still requires validation on independent benchmarks, a systematic hyperparameter study, and a static-baseline comparison -- directions we leave for future work (Section~\ref{conclusion}).

\subsection{Future work}

The approach presents opportunities for expansion and refinement. Future work directions include:
\begin{itemize}
    \item \textbf{Evolving rule discarding}: replacing the total deactivation of a rule with a probabilistic activation rate, making it possible to discover the activation probability that optimizes a rule's own entropy contribution while reducing the traces it collects.
    \item \textbf{State contextualization}: expanding the problem's modeling to consider cluster state (e.g., current CPU usage, time of day) in the decision-making process, transforming the current Bandit formulation into a full Markov Decision Process.
    \item \textbf{Integration with business metrics}: incorporating business-impact metrics into the reward function, aligning the sampling strategy not only with technical diversity but also with business priorities.
    \item \textbf{Validation on independent benchmarks}: extending RADAR's experimental evaluation to industry- and academia-standard reference applications, to test the approach's generalization under denser service topologies, different programming languages, and independent stress workloads.
\end{itemize}

\backmatter





\bmhead{Acknowledgements}

This work was supported by the Coordination for the Improvement of Higher Education Personnel (CAPES) -- Funding Code 001 -- and by the National Council for Scientific and Technological Development (CNPq), PQ Scholarships process no. 308075/2025-0 and 315427/2023-0.

\bibliography{bibliography}

@article{kratzke2018brief,
  title={A brief history of cloud application architectures},
  author={Kratzke, Nane},
  journal={Applied Sciences},
  volume={8},
  number={8},
  pages={1368},
  year={2018},
  publisher={MDPI}
}

@article{zhou2018fault,
  title={Fault analysis and debugging of microservice systems: Industrial survey, benchmark system, and empirical study},
  author={Zhou, Xiang and Peng, Xin and Xie, Tao and Sun, Jun and Ji, Chao and Li, Wenhai and Ding, Dan},
  journal={IEEE Transactions on Software Engineering},
  volume={47},
  number={2},
  pages={243--260},
  year={2018},
  publisher={IEEE}
}

@book{blanco2023practical,
  title={Practical OpenTelemetry: Adopting Open Observability Standards Across Your Organization},
  author={Gomez Blanco, Daniel},
  year={2023},
  publisher={Apress},
  address={Berkeley, CA},
  isbn={978-1-4842-9074-3},
  url={https://link.springer.com/book/10.1007/978-1-4842-9075-0}
}

@inproceedings{hindsight,
  title={The benefit of hindsight: Tracing Edge-Cases in distributed systems},
  author={Zhang, Lei and Xie, Zhiqiang and Anand, Vaastav and Vigfusson, Ymir and Mace, Jonathan},
  booktitle={20th USENIX Symposium on Networked Systems Design and Implementation (NSDI 23)},
  pages={321--339},
  year={2023}
}

@inproceedings{lascasas,
author = {Las-Casas, Pedro and Mace, Jonathan and Guedes, Dorgival and Fonseca, Rodrigo},
title = {Weighted Sampling of Execution Traces: Capturing More Needles and Less Hay},
year = {2018},
isbn = {9781450360111},
address = {New York, NY, USA},
doi = {10.1145/3267809.3267841},
booktitle = {ACM Symposium on Cloud Computing},
pages = {326–332},
numpages = {7},
location = {Carlsbad, CA, USA},
series = {SoCC '18}
}

@article{distributedSystems,
  author       = {Maarten van Steen and Andrew S. Tanenbaum},
  title        = {A brief introduction to distributed systems},
  journal      = {Computing},
  volume       = {98},
  number       = {10},
  pages        = {967--1009},
  year         = {2016},
  doi          = {10.1007/s00607-016-0508-7},
  url          = {https://doi.org/10.1007/s00607-016-0508-7},
  issn         = {1436-5057}
}

@Inbook{Dragoni2017,
author="Dragoni, Nicola
and Giallorenzo, Saverio
and Lafuente, Alberto Lluch
and Mazzara, Manuel
and Montesi, Fabrizio
and Mustafin, Ruslan
and Safina, Larisa",
editor="Mazzara, Manuel
and Meyer, Bertrand",
title="Microservices: Yesterday, Today, and Tomorrow",
bookTitle="Present and Ulterior Software Engineering",
year="2017",
publisher="Springer International Publishing",
address="Cham",
pages="195--216",
isbn="978-3-319-67425-4",
doi="10.1007/978-3-319-67425-4\_12",
url="https://doi.org/10.1007/978-3-319-67425-4\_12"
}

@article{docker_article,
author = {Merkel, Dirk},
title = {Docker: lightweight Linux containers for consistent development and deployment},
year = {2014},
issue_date = {March 2014},
publisher = {Belltown Media},
address = {Houston, TX},
volume = {2014},
number = {239},
issn = {1075-3583},
journal = {Linux J.},
month = mar,
articleno = {2}
}

@book{kubernetes_up_and_running,
  title={Kubernetes: Up and Running: Dive into the Future of Infrastructure},
  author={Burns, B. and Beda, J. and Hightower, K.},
  isbn={9781492046509},
  url={https://books.google.com.br/books?id=-5izDwAAQBAJ},
  year={2019},
  publisher={O'Reilly Media},
  address = {Sebastopol, CA, USA}
}

@misc{jaeger_intro,
  author       = {{Jaeger Authors}},
  title        = {Jaeger Documentation - Version 2.8},
  year         = {2025},
  url          = {https://www.jaegertracing.io/docs/2.8/},
  note         = {Accessed: 2025-08-04}
}

@article{shannon1948mathematical,
  author  = {Shannon, Claude E.},
  title   = {A Mathematical Theory of Communication},
  journal = {Bell System Technical Journal},
  volume  = {27},
  number  = {3},
  pages   = {379--423},
  year    = {1948},
  doi     = {10.1002/j.1538-7305.1948.tb01338.x}
}

@book{cover2006elements,
  author    = {Cover, Thomas M. and Thomas, Joy A.},
  title     = {Elements of Information Theory},
  publisher = {Wiley-Interscience},
  address   = {Hoboken},
  edition   = {2},
  year      = {2006},
  isbn      = {978-0471241959}
}

@book{mackay2003information,
  author    = {MacKay, David J. C.},
  title     = {Information Theory, Inference, and Learning Algorithms},
  publisher = {Cambridge University Press},
  address   = {Cambridge},
  year      = {2003},
  isbn      = {978-0521642989}
}

@Article{xai,
AUTHOR = {Poghosyan, Arnak and Harutyunyan, Ashot and Davtyan, Edgar and Petrosyan, Karen and Baloian, Nelson},
TITLE = {The Diagnosis-Effective Sampling of Application Traces},
JOURNAL = {Applied Sciences},
VOLUME = {14},
YEAR = {2024},
NUMBER = {13},
ARTICLE-NUMBER = {5779},
URL = {https://www.mdpi.com/2076-3417/14/13/5779},
ISSN = {2076-3417},
DOI = {10.3390/app14135779}
}

@inproceedings{audit,
author = {Luo, Liang and Nath, Suman and Sivalingam, Lenin Ravindranath and Musuvathi, Madan and Ceze, Luis},
title = {Troubleshooting transiently-recurring problems in production systems with blame-proportional logging},
year = {2018},
isbn = {9781931971447},
publisher = {USENIX Association},
address = {USA},
booktitle = {Proceedings of the 2018 USENIX Conference on Usenix Annual Technical Conference},
pages = {321–334},
numpages = {14},
location = {Boston, MA, USA},
series = {USENIX ATC '18}
}

@INPROCEEDINGS{ebpf-enhanced,
  author={Sharma, Bhavye and Nadig, Deepak},
  booktitle={ICC 2024 - IEEE International Conference on Communications}, 
  title={{eBPF-Enhanced Complete Observability Solution for Cloud-native Microservices}}, 
  year={2024},
  volume={},
  number={},
  pages={1980-1985},
  doi={10.1109/ICC51166.2024.10622329}}

@inproceedings{hubble,
author = {Yu Luo and Kirk Rodrigues and Cuiqin Li and Feng Zhang and Lijin Jiang and Bing Xia and David Lion and Ding Yuan},
title = {Hubble: Performance Debugging with {In-Production}, {Just-In-Time} Method Tracing on Android},
booktitle = {16th USENIX Symposium on Operating Systems Design and Implementation (OSDI 22)},
year = {2022},
isbn = {978-1-939133-28-1},
address = {Carlsbad, CA},
pages = {787--803},
url = {https://www.usenix.org/conference/osdi22/presentation/luo},
publisher = {USENIX Association},
month = jul
}

@book{ObservEng,
  title={Observability Engineering},
  author={Majors, C. and Fong-Jones, L. and Miranda, G.},
  isbn={9781492076414},
  url={https://books.google.com.br/books?id=KGZuEAAAQBAJ},
  year={2022},
  pages={320},
  edition = {1st},
  address = {Sebastopol, CA, USA},
  publisher={O'Reilly Media}
}

@book{sutton2018reinforcement,
  author    = {Sutton, Richard S. and Barto, Andrew G.},
  title     = {Reinforcement Learning: An Introduction},
  publisher = {MIT Press},
  address   = {Cambridge, MA},
  edition   = {2nd},
  year      = {2018},
  isbn      = {978-0262039246}
}

@article{kephart2003vision,
  author  = {Kephart, Jeffrey O. and Chess, David M.},
  title   = {The Vision of Autonomic Computing},
  journal = {Computer},
  volume  = {36},
  number  = {1},
  pages   = {41--50},
  year    = {2003},
  doi     = {10.1109/MC.2003.1160055}
}

@article{yang2024micronet,
  author  = {Yang, Jingjing and Guo, Yuchun and Chen, Yishuai and Zhao, Yongxiang},
  title   = {MicroNet: Operation Aware Root Cause Identification of Microservice System Anomalies},
  journal = {IEEE Transactions on Network and Service Management},
  volume  = {21},
  number  = {4},
  pages   = {4255--4267},
  year    = {2024},
  doi     = {10.1109/TNSM.2024.3387552}
}

@article{li2025tracedae,
  author  = {Li, Junjun and Ying, Shi and Li, Tiangang and Tian, Xiangbo},
  title   = {TraceDAE: Trace-Based Anomaly Detection in Microservice Systems via Dual Autoencoder},
  journal = {IEEE Transactions on Network and Service Management},
  volume  = {22},
  number  = {5},
  pages   = {4884--4897},
  year    = {2025},
  doi     = {10.1109/TNSM.2025.3583213}
}

@article{samani2024servicemesh,
  author  = {Samani, Forough Shahab and Stadler, Rolf},
  title   = {A Framework for Dynamically Meeting Performance Objectives on a Service Mesh},
  journal = {IEEE Transactions on Network and Service Management},
  volume  = {21},
  number  = {6},
  pages   = {5992--6007},
  year    = {2024},
  doi     = {10.1109/TNSM.2024.3434328}
}

@inproceedings{shaikh2025autoscaling,
  author    = {Shaikh, Faraz and Reali, Gianluca and Femminella, Mauro},
  title     = {Intelligent Autoscaling with Attention-based Reinforcement Learning for SLA-Aware Resource Management in Edge-Cloud Environments},
  booktitle = {2025 21st International Conference on Network and Service Management (CNSM)},
  year      = {2025},
  doi       = {10.23919/CNSM67658.2025.11297411}}

\end{document}